\documentclass[11pt,fullpage]{article}

\newcommand{\ol}{\setlength{\itemsep}{0pt.}\begin{enumerate}}
\newcommand{\eol}{\end{enumerate}\setlength{\itemsep}{-\parsep}}
\newcommand{\ignore}[1]{}
\begin{document}
\begin{center}
{\bf Classical deterministic complexity of Edmonds' problem and
 Quantum Entanglement
 }

\vskip 4pt
{Leonid Gurvits }\\
\vskip 4pt
{\tt gurvits@lanl.gov}\\
\vskip 4pt
Los Alamos National Laboratory, Los Alamos , NM 87545 , USA.
\end{center}




\begin{abstract}
Generalizing a decision problem for bipartite perfect matching ,
J. Edmonds introduced in \cite{edmonds67} the problem (now 
known as the Edmonds Problem) 
of deciding if a given linear subspace of $M(N)$
contains a nonsingular matrix, where $M(N)$ stands for the linear space
of complex $N \times N$ matrices.  This problem led to many
fundamental developments in matroid theory etc. \\ Classical matching
theory can be defined in terms of matrices with nonnegative entries.
The notion of Positive operator, central in Quantum Theory, is a
natural generalization of matrices with nonnegative entries. 
(Here operator refers to maps from matrices to matrices.) First,
we reformulate the Edmonds Problem in terms of of completely positive
operators, or equivalently, in terms of bipartite density matrices .
It turns out that one of the most important cases when Edmonds' problem
can be solved in polynomial deterministic time, i.e. an intersection
of two geometric matroids, corresponds to unentangled (aka separable
) bipartite density matrices .  We introduce a very general class (or
promise ) of linear subspaces of $M(N)$ on which there exists a
polynomial deterministic time algorithm to solve Edmonds' problem . \\
The algorithm is a thoroughgoing generalization of algorithms in
\cite{lsw}, \cite{GY}, and its analysis benefits from an operator
analog of permanents, so called Quantum Permanents .  Finally, we
prove that the weak membership problem for the convex set of separable
normalized bipartite density matrices is NP-HARD. \\

\end{abstract} 
 
 
\newtheorem{THEOREM}{Theorem}[section]
\newenvironment{theorem}{\begin{THEOREM} \hspace{-.85em} {\bf :} 
}%
                        {\end{THEOREM}}
\newtheorem{LEMMA}[THEOREM]{Lemma}
\newenvironment{lemma}{\begin{LEMMA} \hspace{-.85em} {\bf :} }%
                      {\end{LEMMA}}
\newtheorem{COROLLARY}[THEOREM]{Corollary}
\newenvironment{corollary}{\begin{COROLLARY} \hspace{-.85em} {\bf 
:} }%
                          {\end{COROLLARY}}
\newtheorem{PROPOSITION}[THEOREM]{Proposition}
\newenvironment{proposition}{\begin{PROPOSITION} \hspace{-.85em} 
{\bf :} }%
                            {\end{PROPOSITION}}
\newtheorem{DEFINITION}[THEOREM]{Definition}
\newenvironment{definition}{\begin{DEFINITION} \hspace{-.85em} {\bf 
:} \rm}%
                            {\end{DEFINITION}}
\newtheorem{EXAMPLE}[THEOREM]{Example}
\newenvironment{example}{\begin{EXAMPLE} \hspace{-.85em} {\bf :} 
\rm}%
                            {\end{EXAMPLE}}
\newtheorem{CONJECTURE}[THEOREM]{Conjecture}
\newenvironment{conjecture}{\begin{CONJECTURE} \hspace{-.85em} 
{\bf :} \rm}%
                            {\end{CONJECTURE}}
\newtheorem{PROBLEM}[THEOREM]{Problem}
\newenvironment{problem}{\begin{PROBLEM} \hspace{-.85em} {\bf :} 
\rm}%
                            {\end{PROBLEM}}
\newtheorem{QUESTION}[THEOREM]{Question}
\newenvironment{question}{\begin{QUESTION} \hspace{-.85em} {\bf :} 
\rm}%
                            {\end{QUESTION}}
\newtheorem{REMARK}[THEOREM]{Remark}
\newenvironment{remark}{\begin{REMARK} \hspace{-.85em} {\bf :} 
\rm}%
                            {\end{REMARK}}
 
\newcommand{\thm}{\begin{theorem}}
\newcommand{\lem}{\begin{lemma}}
\newcommand{\pro}{\begin{proposition}}
\newcommand{\dfn}{\begin{definition}}
\newcommand{\rem}{\begin{remark}}
\newcommand{\xam}{\begin{example}}
\newcommand{\cnj}{\begin{conjecture}}
\newcommand{\prb}{\begin{problem}}
\newcommand{\que}{\begin{question}}
\newcommand{\cor}{\begin{corollary}}
\newcommand{\prf}{\noindent{\bf Proof:} }
\newcommand{\ethm}{\end{theorem}}
\newcommand{\elem}{\end{lemma}}
\newcommand{\epro}{\end{proposition}}
\newcommand{\edfn}{\bbox\end{definition}}
\newcommand{\erem}{\bbox\end{remark}}
\newcommand{\exam}{\bbox\end{example}}
\newcommand{\ecnj}{\bbox\end{conjecture}}
\newcommand{\eprb}{\bbox\end{problem}}
\newcommand{\eque}{\bbox\end{question}}
\newcommand{\ecor}{\end{corollary}}
\newcommand{\eprf}{\bbox}
\newcommand{\beqn}{\begin{equation}}
\newcommand{\eeqn}{\end{equation}}
\newcommand{\wbox}{\mbox{$\sqcap$\llap{$\sqcup$}}}
\newcommand{\bbox}{\vrule height7pt width4pt depth1pt}
\newcommand{\qed}{\bbox}
\def\sup{^}
\def\Tp{Tchebyshef polynomial}
\def\Tps{TchebysDeto be the maximafine $A(n,d)$ l size of a code with distance 
$d$hef polynomials}
\newcommand{\rarrow}{\rightarrow}
\newcommand{\larrow}{\leftarrow}
\newcommand{\grad}{\bigtriangledown}

\overfullrule=0pt
\def\setof#1{\lbrace #1 \rbrace}
\section{ Introduction and Main Definitions}
Let $M(N)$ be the linear space of $N \times N$ complex matrices . The following fundamental problem has been posed by  J. Edmonds
in \cite{edmonds67}: 
\prb
 Given a linear subspace $V \subset M(N)$  to decide if there exists a nonsingular matrix $A \in V$ .   
\eprb
We will assume throughout the paper that the subspace $V$ is presented as a finite spanning $k$-tuple of rational matrices
$S(V) = \{A_1,...,A_k\} ( k \leq N^{2} )$, i.e .  the 
linear space generated by them is equal to $V$. As usual, the complexity
parameter of the input $<S(V)>$  is equal to ( $N$ + ``number of bits of entries of matrices $A_i, 1 \leq i \leq k $'' ). \\
Thus Edmonds' problem is equivalent to checking if the following determinantal polynomial
$$
P_{{\bf A}}(x_1,...,x_k) = \det( \sum_{ 1 \leq i \leq k } x_{i}A_{i} )
$$
is not identically equal to zero. \\
This determinantal polynomial can be efficiently evaluated, hence randomized 
poly-time  algorithms, based on Schwartz's lemma or its recent improvements, are readily
available (notice that our problem is defined over infinite field with infinite characteristic). \\
But for general linear subspaces of M(N), i.e. without an extra assumption 
(promise),
 poly-time deterministic algorithms are not known and the problem is believed to be "HARD" . \\
Like any other homogeneous polynomial, $P_{{\bf A}}(x_1,...,x_k)$  is a weighted sum of monomials of degree $N$, i.e.
 \beqn
P_{{\bf A}}(x_1,...,x_k)  =  \sum_{(r_1,...,r_k) \in I_{k,N}} a_{r_1,...,r_k} x_{1}^{r_1} x_{2}^{r_2}...x_{k}^{r_k},
\eeqn
where $I_{k,N}$ stands for a set of vectors $r = (r_1,...,r_k)$   with nonnegative integer components and  $\sum_{1 \leq i \leq k} r_i = N$ . \\
We will make substantial use of the following (Hilbert) norm of determinantal polynomial $P(.)$ :
\beqn
\|P\|_{G}^{2} = :  \sum_{(r_1,...,r_k) \in I_{k,N}} |a_{r_1,...,r_k}|^{2} r_{1}! r_{2}! ...r_{k}! 
\eeqn
It is easy to show that the determinantal polynomial $P_{{\bf A}}(x_1,...,x_k) \equiv 0$ iff $P_{{\bf A}}(r_1,...,r_k) = 0$ for
all $(r_1,...,r_k) \in I_{k,N}$, which amounts to $|I_{k,N}| = \frac{(N+k-1)!}{N! (k-1)!}$  computations of determinants . \\
We will show that $\|P\|_{G}^{2}$ can be evaluated in $O(2^{N}N!)$  computations of determinants   . \\
More importantly,  $\|P\|_{G}^{2}$  serves as a natural tool to analyze our main algorithm . \\
The algorithm to solve Edmonds' problem, which we  introduce and analyze later in the paper, is a rather thoroughgoing generalization
of  the recent algorithms \cite{lsw}, \cite{GY} for deciding the existence of perfect matchings. They are based on
so-called Sinkhorn's iterative scaling. The algorithm in \cite{GY} is a greedy version of Sinkhorn's scaling and has been 
analyzed using KLD-divergence; the algorithm in \cite{lsw} is a standard Sinkhorn's scaling and a "potential"
used for its analysis is the permanent.  Our analysis is a sort of combination of techniques from \cite{lsw}, \cite{GY} .
Most importantly, $\|P\|_{G}^{2}$  can be viewed as a generalization of the permanent. \\
The organization of this paper proceeds as follows. In Section 2 we will recall fundamental notions from
Quantum Information Theory such as bipartite density matrix, positive and completely positive operator,
separability and entanglement.  After that we will rephrase Edmonds' problem using those notions and reformulate
the famous Edmonds-Rado theorem on the rank of intersection of two geometric matroids in terms of the rank non-decreasing
property of the corresponding
(separable) completely positive operator.  We will end Section 2  by introducing a property, called the Edmonds-Rado property,
of linear subspaces  of $M(N)$ which
allows a poly-time deterministic algorithm to solve Edmonds' problem and will explain how is this property is
related to quantum entanglement. \\
In Section 3 we will express $G$-norm of a determinantal polynomial $P_{{\bf A}}(x_1,...,x_k) $ in terms of
the associated bipartite density matrix, and we 
will prove various inequalities and properties of $G$-norm which will
be needed later on for the analysis of the main algorithm. \\
In Section 4 we will introduce and analyze the main algorithm of the paper, Operator Sinkhorn Scaling. \\
In Section 5 we will apply this algorithm to solve Edmonds' problem for linear subspaces  of $M(N)$ having the Edmonds-Rado property.
In Section 6 we will prove NP-HARDNESS of the weak membership problem for the compact convex set
of separable normalized density matrices.
Finally, in the Conclusion section we will pose several open problems and  directions for future research. \\
We would like to stress that our paper does not contain explicit connections to Quantum Computing.
It rather aims to study quantum entanglement from the point of view of classical computational
complexity and computational geometry  and to use some ideas and structures from Quantum Information Theory
to construct and analyse classical algorithms. \\
The main algorithm of this paper is a third "generation" of scalings applications to computer science problems ,
starting with ( \cite{lsw} , \cite{GY} ; applied to bipartite perfect matchings and an approximation of the permanent ) and
(\cite{GS} , \cite{GS1} ; applied to an approximation of the mixed discriminant and mixed volume ) . \\
And here it is used to solve very non-trivial , important and seemingly different problem .
\section{Bipartite density matrices, completely positive operators and Edmonds Problem }
\dfn
A positive semidefinite matrix $\rho_{A,B} : C^{N} \otimes C^{N} \rightarrow C^{N} \otimes C^{N} $
is called a bipartite unnormalized density matrix \\
 ({\bf BUDM }).  If $tr(\rho_{A,B}) =1$ then this $\rho_{A,B}$ 
is called a bipartite density matrix. \\
It is convenient to represent a bipartite $\rho_{A,B} = \rho(i_{1},i_{2},j_{1},j_{2})$ as
the following block matrix :
\beqn 
\rho_{A,B} = \left( \begin{array}{cccc}
		  A_{1,1} & A_{1,2} & \dots & A_{1,N}\\
		  A_{2,1} & A_{2,2} & \dots & A_{2,N}\\
		  \dots &\dots & \dots & \dots \\
		  A_{N,1} & A_{N,2} & \dots & A_{N,N}\end{array} \right),
\eeqn
where $A_{i_{1},j_{1}} = : \{ \rho(i_{1},i_{2},j_{1},j_{2}) : 1 \leq i_{2},j_{2} \leq N \},
 1 \leq i_{1},j_{1} \leq N $. \\
A ({\bf BUDM })   $\rho$ is called {\bf separable } if
\beqn
\rho = \rho_{(X,Y)} = : \sum_{1 \leq i \leq K} x_{i}x_{i}^{\dagger} \otimes y_{i}y_{i}^{\dagger},
\eeqn
and  {\bf entangled } otherwise. \\
If the vectors $ x_{i}, y_{i} ; 1 \leq i \leq K $ in (6) are real then $\rho$ is is called {\bf real separable }. \\
The quantum marginals are defined as $\rho_{A} = \sum_{1 \leq i \leq N}A_{i,i} $ and  \\
$ \rho_{B}(i,j) = tr(A_{i,j} ) ;   1 \leq i, j \leq N $.\\
Next we define  the ({\bf BUDM })   $\rho_{{\bf A}}  $   associated with the
$k$-tuple \\ 
$ {\bf A} = (A_1,...,A_k) $:
\beqn
\rho_{{\bf A}}(i_{1},i_{2},j_{1},j_{2}) = : \sum _{1 \leq l \leq k}  A_{l}(i_{1},i_{2}) \overline{A_{l}(j_{1},j_{2}) },
\eeqn
where for a complex number $z= x + iy$ its conjugate $\bar{z} = x - iy$. \\
Rewriting expression (5) in  terms of blocks of $\rho_{{\bf A}}$ as in (3), we get that
$$
 A_{i,j} = \sum _{1 \leq l \leq k} A_{l}e_{i}e_{j}^{\dagger}A_{l}^{\dagger},  1 \leq i,j \leq N.
 $$
(In quantum physics language, one can view a tuple $ {\bf A} = (A_1,...,A_k) $ of complex matrices as a tuple
of unnormalized bipartite "wave functions" ; and ({\bf BUDM })   $\rho_{{\bf A}}  $  as a corresponding mixed bipartite state. ) \\
We will call ({\bf BUDM }) $\rho$ weakly separable
if there exists a separable $\rho'_{(X,Y)}$ with the same image as  $\rho$: $Im(\rho)= Im(\rho'_{(X,Y)})$. \\
( Recall that in this finite dimensional case $Im(\rho)$ is the linear subspace formed by all linear combinations
of columns of matrix $\rho$.) \\
A linear operator $T: M(N) \rightarrow M(N)$ is called positive if $T(X) \succeq 0$ for all $X  \succeq 0$,
and strictly positive if $T(X) \succeq \alpha tr(X)I$ for all $X  \succeq 0$  and some $\alpha > 0$.
A positive operator T is called completely  positive if
\beqn 
T(X) = \sum_{1 \leq i \leq N^{2}} A_{i} X  A_{i}^{\dagger} ;     A_{i},  X  \in M(N)
\eeqn
Choi's representation of the linear operator $T: M(N) \rightarrow M(N)$  is a block matrix  $CH(T)_{i,j} = : T(e_{i}e_{j}^{\dagger}).$
The dual to $T$ with respect to the inner product $<X,Y> =tr(XY^{\dagger}) $ is denoted as $T^{*}$.
A very useful and easy result of Choi states that $T$ is completely  positive iff $CH(T)$ is   ({\bf BUDM }).
Using this natural (linear) correspondence between completely  positive operators and ({\bf BUDM }), we will
freely "transfer" properties of ({\bf BUDM }) to completely  positive operators. For example,
a linear operator $T$ is called separable iff $CH(T)$ is separable, i.e.
\beqn
T(Z) = T_{(X,Y)}(Z) = \sum_{1 \leq i \leq K} x_{i}y_{i}^{\dagger}  Z  y_{i}x_{i}^{\dagger}
\eeqn
Notice that $CH(T_{(X,Y)})= \rho_{(\bar{Y},X)}$  and  $T_{(X,Y)}^{*}= T_{(Y,X)}$. \\
(The components of the vector $\bar{y}$ are the
complex conjugates of corresponding components of $y$ ).
\edfn
\rem
There is a natural (column by column ) correspondence between $M(N)$ and  $C^{N^{2}} \cong C^{N} \otimes C^{N}$. It works as follows
\begin{eqnarray*}
&\{A(i,j), 1 \leq i,j \leq N \} \in M(N)  \Leftrightarrow  \\  
&(A(1,1),...,A(1,N);......;A(1,N),...,A(N,N))^{T} \in C^{N^{2}}
\end{eqnarray*}
In light of definition (2.1), we will represent a linear subspace $ V \subset M(N) \cong C^{N} \otimes C^{N}$ in  Edmonds Problem
as the image of the
 ({\bf BUDM })  $\rho$. And as the complexity measure we will use the number
of bits of (rational) entries of  $\rho$ plus the dimension $N$.
\erem

\dfn
A positive linear operator $T:M(N) \rightarrow M(N)$ is called rank non-decreasing iff
\beqn
Rank(T(X)) \geq Rank (X) \mbox{ if  } X \succeq 0 ;
\eeqn
and is called indecomposable iff
\beqn
Rank(T(X)) > Rank(X) \mbox{ if} X \succeq 0 \mbox{ and} 1 \leq Rank(X) < N.
\eeqn
A positive linear operator $T:M(N) \rightarrow M(N)$ is called doubly stochastic iff $T(I)=I$ and $T^{*}(I)=I$ ;
called $\epsilon$ - doubly stochastic iff $DS(T) =:  tr((T(I)-I)^{2}) + tr((T^{*}(I)-I)^{2}) \leq \epsilon^{2} $.
\edfn

The next Proposition(2.4) is a slight generalization of the corresponding result in \cite{lsw}.
\pro
Doubly stochastic operators are rank non-decreasing. If either $T(I)=I$ or $T^{*}(I)=I$   and   $DS(T) \leq N^{-1}$
then $T$ is rank non-decreasing. If  $DS(T) \leq (2N+1)^{-1}$  then $T$ is rank non-decreasing. 
\epro
Let us consider a completely positive operator $ T_{{\bf A}} : M(N) \rightarrow  M(N), \\ T(X) = \sum_{1 \leq i \leq k} A_{i} X  A_{i}^{\dagger}$, 
and let $L(A_{1},A_{2},...,A_{K})$ be a linear subspace of $M(N)$ generated by matrices $\{A_{i}, 1 \leq i \leq k \}$.
It is easy to see that if $\hat{A} \in  L(A_{1},A_{2},...,A_{k})$ then $\hat{A}(Im(X)) \subset Im(T(X))$ for all   $X \succeq 0$. \\
Therefore, if $L(A_{1},A_{2},...,A_{k})$ contains a nonsingular matrix then the operator $T$ is rank non-decreasing. \\
This simple observation suggested the following property of linear subspaces of $M(N)$ : \\
{\bf Edmonds-Rado Property (ERP) } : \\
 A linear subspace $V=L(A_{1},A_{2},...,A_{k})$ has the {\bf (ERP) } property if
the existence of nonsingular matrix in $V$ is equivalent to the fact that  
the  associated completely positive operator $T_{{\bf A}}$ 
is rank  non-decreasing.  In other words, a  linear subspace $V \subset M(N)$ has the {\bf (ERP) } property if the fact that all
matrices in $V$ are singular is  equivalent to the existence of two linear subspaces $X,Y \subset C^{N}$ such
$dim(Y) < dim(X)$ and  $A(X) \subset Y$  for all matrices $A \in V$. \\ 
The main "constructive" result of this paper is that for linear subspaces of $M(N)$ having the {\bf ERP } there
is a deterministic poly-time algorithm to solve Edmonds' problem. \\
In the rest of this section we will explain why we chose to call this property Edmonds-Rado, will describe a rather wide
class of linear subspaces with {\bf (ERP) } property and will give an example of a subspace without it.\\

\subsection{ Examples of linear subspaces of $M(N)$ having {\bf Edmonds-Rado Property  } }
Let us first list some obvious but useful facts about the Edmonds-Rado property.
\begin{enumerate}
\item
Suppose that $V=L(A_{1},A_{2},...,A_{k}) \subset M(N) $ has the {\bf (ERP) }  and $C,D \in M(N)  $ are two nonsingular matrices.
Then linear subspace $V_{C,D} = : L(CA_{1}D, CA_{2}D,..., CA_{k}D) $  also has the {\bf (ERP) }.
\item
If $V=L(A_{1},A_{2},...,A_{k}) \subset M(N) $ has the {\bf (ERP) } then both
$V^{\dagger}= : L(A_{1}^{\dagger},A_{2}^{\dagger},...,A_{k}^{\dagger}) $ and \\
$V^{T}=L(A_{1}^{T},A_{2}^{T},...,A_{k}^{T})$ have the {\bf (ERP) }.
\item Any linear subspace $V=L(A_{1},A_{2},...,A_{k}) \subset M(N) $ with matrices $\{A_{i}, 1 \leq i \leq k \}$ being 
positive semidefinite  has the {\bf (ERP)}.
\item
Suppose that linear  subspaces $V=L(A_{1},A_{2},...,A_{k}) \subset M(N_{1})$ and $W= L(B_{1},B_{2},...,B_{k}) \subset M(N_{2})$
both have the {\bf (ERP)}. 
Define the following matrices $C_{i} \in M(N_{1}+ N_{2}), 1 \leq i \leq k $ :
$$
 C_{i} = \left( \begin{array}{cc}
		  A_{i}& D_{i}\\
		  0& B_{i} \end{array} \right) 	  
$$
Then the linear subspace $L(C_{1},C_{2},...,C_{k}) \subset M(N_{1}+ N_{2}) $ also has the {\bf (ERP) }. \\
A particular case of this fact is that any linear subspace of $M(N)$ which has a basis consisting of upper diagonal
matrices has the {\bf (ERP)}.
\item
Any 1-dimensional subspace of $M(N)$ has the {\bf (ERP) } property.
\end{enumerate}
The next theorem gives the most interesting example which motivated the name "{\bf Edmonds-Rado Property }".
Let us first recall one of the most fundamental results in matroids theory, i.e. the Edmonds-Rado characterization of the rank of the
intersection of two geometric matroids.
\dfn
The intersection of two geometric matroids
$MI(X,Y) = \{ (x_{i},y_{i}), 1 \leq i \leq K \}$ is a finite family of distinct $2$-tuples of non-zero
$N$-dimensional complex vectors, i.e. $x_{i},y_{i} \in C^{N}$. \\
The rank of $MI(X,Y)$, denoted by $Rank( MI(X,Y)$ is the largest integer $m$ such that there exist $1 \leq i_{1} <...<i_{m} \leq K $
with both sets $\{x_{i_{1}},...,x_{i_{m}} \}$ and $\{y_{i_{1}},...,y_{i_{m}} \}$ being
linearly independent. 
\edfn
The Edmonds-Rado theorem (  \cite{gr:lo:sc} ) states   (in the 
much more general situation of the 
intersection of any two matroids with a common ground set)
that 
\begin{eqnarray}
&Rank(MI(X,Y)) = \\
&\min_{S \subset \{1,2,...,K\}} dimL(x_{i} ; i \in S) + dimL(y_{j}; j \in \bar{S} )
\end{eqnarray}
It is easy to see that $Rank(MI(X,Y))$ is the maximum rank achieved in 
the linear subspace $L(x_{1}y_{1}^{\dagger},...,x_{K}y_{K}^{\dagger} )$ ;
and $Rank(MI(X,Y)) = N$ iff $L(x_{1}y_{1}^{\dagger},...,x_{K}y_{K}^{\dagger} )$ contains a nonsingular matrix.
\thm
Suppose that $T : M(N) \rightarrow M(N)$, $T(X)= \sum_{1 \leq j \leq l} A_{i}X A_{i}^{\dagger}$, is a completely positive weakly separable operator,
i.e. there exists a  family of rank one matrices $\{x_{1}y_{1}^{\dagger},...,x_{l}y_{l}^{\dagger} \} \subset M(N)$
such that $L(A_{1},...,A_{L}) = L(x_{1}y_{1}^{\dagger},...,x_{l}y_{l}^{\dagger})$.

Then the following conditions are equivalent :
\begin{description}
\item [Fact 1] 
$T$ is rank non-decreasing.
\item [Fact 2]
The rank of intersection of two geometric matroids $MI(X,Y)$ is equal to $N$.
\item [Fact 3]
The exists a nonsingular matrix $A$ such that $ Im(AXA^{\dagger}) \subset Im(T(X)), X \succeq 0 $.
\item [Fact 4]
The exists a nonsingular matrix $A$ such that the operator $T'(X)= T(X) - AXA^{\dagger} $ is completely positive.
\end{description}
\ethm
\prf  $ [ 2 \Longrightarrow 1  ]$ \\
Suppose that the rank of $MI(X,Y)$ is equal to $N$. Then 
$$ 
RankT(X) = dim(L(x_{i}  ;  i \in S )) \mbox{ where }  S = : \{ i :   y_{i}^{\dagger} X y_{i} \neq 0 \}
$$
As $dim (L(y_{j}  ; j \in \bar{S} ) \leq \ dim (Ker(X)) = N - Rank(X)$ hence, from the Edmonds-Rado Theorem 
we get that \\
 $Rank(T(X))  \geq N - (N - Rank(X)) = Rank(X) $. \\
$[1  \Longrightarrow 2 ]$   Suppose that $T$ is rank non-decreasing and for any $S \subset  \{1,2,...,l\} $ consider an orthogonal
projector $P \succeq 0$ on \\
 $L(y_{j}  ; j \in \bar{S} )^{\perp}$. Then
\begin{eqnarray*}
&dim(L(x_{i} :   i \in S ) )  \geq  \\
&Rank(T(P) ) \geq Rank(P) = N -dim (L(y_{j}  ; j \in \bar{S} ) ).
\end{eqnarray*}
It follows from the Edmonds-Rado Theorem that the 
rank of  $MI(X,Y)$ is equal to $N$. 
All other "equivalences" follow now directly. \\
\eprf
\rem
 Theorem 2.6  makes  the Edmonds-Rado theorem sound like  Hall's theorem on bipartite perfect matchings.\\
 Indeed,  consider a weighted  incidence matrix $A_{\Gamma}$ of a
bipartite graph $\Gamma$, i.e.$ A_{\Gamma}(i,j) > 0$  if  $i$ from the first part is adjacent to
$j$ from the second part and equal to zero otherwise. Then Hall's theorem can be immediately
reformulated as follows : \\
A perfect matching, which is just a permutation in this bipartite case, exists iff \\
$|A_{\Gamma} x|_{+} \geq |x|_{+}$ for any vector $x$ with nonnegative
entries, where $|x|_{+}$ stands for the number of positive entries of a vector $x$. \\
All known algorithms (for instance, linear programming based  on 
\cite{gr:lo:sc}) to
compute the rank of the 
intersection of two geometric matroids  require an explicit knowledge
of pairs of vectors $(x_{i},y_{i})$, or, in other words, an explicit representation of the
rank one basis $ \{x_{i} y_{i}^{\dagger}, 1 \leq i \leq l \} $. The algorithm in this
paper requires only a promise that such a rank one basis (not necessarily rational!) does exist.
\erem
Another example comes from \cite{marek}.  Consider pairs of  matrices $ (A_{i},B_{i} \in M(N) ; 1 \leq i \leq K$. \\
Let $V_{i} \subset M(N)$ be the linear subspace of all matrix solutions of the equation $XA_{i}=B_{i}X$. \\
One of the problems solved in  \cite{marek} is to decide if $W = V_{1} \cap... \cap V_{K}$ contains a nonsingular matrix.\\
It is not clear to the author whether the class of such linear subspaces $W$ satisfies  the {\bf (ERP) } property. \\
But suppose that $A_{1}$ is similar to $B_{1}$ ($V_{1}$ contains a nonsingular matrix ) and, additionally,
assume that $dim(Ker(A_{1}- \lambda I) = dim(Ker(B_{1}- \lambda I)  \leq 1$ for all complex $\lambda \in C$. \\
(I.e.  just one Jordan block for each eigenvalue.) \\
It is not difficult to show that in this case there exist two nonsingular matrices $D, Q$  and upper diagonal
matrices $(U_{1},...,U_{r})$ such that $V_{1} = L(DU_{1}Q,..., DU_{r}Q)$. It follows, using \\
Facts (1, 4 ) above,
that $V_{1}$ as well
as any of its linear subspaces satisfy {\bf (ERP)}. \\
\xam
Consider the following completely positive doubly stochastic operator $Sk_{3} : M(3) \rightarrow M(3)$ :
\beqn
Sk_{3}(X) = \frac{1}{2} ( A_{(1,2)}XA_{(1,2)}^{\dagger} + A_{(1,3)}XA_{(1,3)}^{\dagger} + A_{(2,3)}XA_{(2,3)}^{\dagger} )
\eeqn
Here $\{ A_{(i,j)}, 1 \leq i < j \leq 3 \}$ is a standard basis in the linear
subspace $K(3) \subset M(3)$
consisting of all skew-symmetric matrices, i.e. 
$A_{(i,j)} = : e_{i} e_{j}^{\dagger} - e_{j}e_{i}^{\dagger}$
and $\{ e_{i}, 1 \leq i \leq 3 \}$ is a standard orthonormal basis in $C^{3}$. \\
It is clear that all $3 \times 3$ skew-symmetric matrices are singular. As $Sk_{3}$ is a completely positive doubly stochastic operator,
and, thus, is rank non-decreasing, therefore $K(3) \subset M(3)$ is an example of a linear subspace not having {\bf (ERP) } property. \\
More "exotic" properties of this operator can be found in \cite{quant}.
\exam
\section{Quantum permanents and $G$-norms of determinantal polynomials}
Consider a $k$-tuple of  $N \times N$ complex matrices $ {\bf A} = (A_1,...,A_k) $.
Our first goal here is to express the square of the $G$-norm of a determinantal
polynomial $P_{{\bf A}}(x_1,...,x_k)$ in terms of the associated bipartite density matrix  ({\bf BUDM })   $\rho_{{\bf A}}$,
which is defined as in (5). \\
Consider an $N$-tuple of complex $N \times N$ matrices, ${\bf B} = (B_1,...,B_N)$.
Recall that the mixed discriminant  $M({\bf B}) = M(B_1,...,B_N)$ is defined as follows :
\beqn 
M(B_1,...B_N) =   \frac{\partial^n}{\partial x_1
... \partial x_N}  \det(x_1 B_1 +....+x_N B_N).
\eeqn
Or  equivalently :
\beqn
M(B_1,...B_N) =  \sum_{\sigma,\tau \in S_N}
(-1)^{sign(\sigma\tau)} \prod_{i=1}^N B_i(\sigma(i), \tau(i)),
\eeqn
where $S_{n}$ is the symmetric group, i.e. the group of all permutations of the set $\{1,2, \cdots, N\}$. 
If matrices $B_{i}, 1 \leq i \leq N $ are diagonal then their mixed discriminant is equal
to the corresponding permanent (\cite{GS}). 
\dfn
Let us consider a block matrix $\rho$ as in (3) (not necessarily positive semidefinite).
We define the quantum permanent,
$ QP( \rho )$, by the following equivalent formulas : 
\beqn
QP( \rho ) = :  \sum_{\sigma \in S_N} (-1)^{sign(\sigma)}M(A_{1,\sigma(1)},...,A_{N,\sigma(N)}) ;
\eeqn

\begin{eqnarray}
QP( \rho ) & = & \frac{1}{N!}\sum_{\tau_{1}, \tau_{2}, \tau_{3}, \tau_{4} \in S_N}(-1)^{sign(\tau_{1}\tau_{2}\tau_{3}\tau_{4})} \nonumber \\
   & &\prod_{i=1}^N rho(\tau_{1}(i),\tau_{2}(i),\tau_{3}(i), \tau_{4}(i)).  
\end{eqnarray}
\edfn
Straight from this definition, we get the following inner product formula for quantum permanents :
\beqn
QP( \rho ) = <\rho^{\otimes N} Z, Z > , 
\eeqn
where $\rho^{\otimes N}$ stands for a tensor product of
$N$ copies of $\rho$, $<.,. >$ is a standard inner product and \\
$Z(j_{1}^{(1)},j_{2}^{(1)};...;j_{1}^{(N)},j_{2}^{(N)}) = \frac{1}{N!^{\frac{1}{2}}} (-1)^{sign(\tau_{1} \tau_{2})} $  \\
if $ j_{k}^{(i)} = \tau_{k}(i) ( 1 \leq i \leq N ) ;   \tau_{k} \in S_N  (k= 1,2) $ and zero otherwise. \\
\rem
Notice that the equality (17) implies that if \\
$ \rho_{1} \succeq  \rho_{2} \succeq 0$ then 
$QP( \rho _{1}) \geq QP( \rho _{2}) \geq 0$ . \\
The standard norm of $N^{2N}$-dimensional vector $Z$ defined above is
equal to 1. Thus, if $\rho$ is a normalized bipartite density matrix then $QP( \rho )$ can be viewed as the probability
of a particular outcome of some (von Neumann) measurement. Unfortunately, in this case $QP( \rho ) \leq \frac{N!}{N^{N}}$. \\
Consider an arbitrary permutation $\sigma \in S_{4}$ and for a block matrix (or tensor ) $\rho = \{\rho(i_1,i_2,i_3,i_4) ;  
1 \leq i_1,i_2,i_3,i_4 \leq N \}$ define $\rho^{\sigma} = \{\rho(i_{\sigma(1)},i_{\sigma(2)},i_{\sigma(3)},i_{\sigma(4)} \} $.
It is easy to see that $QP( \rho ) = QP( \rho^{\sigma} ) $.  Another simple but important fact about quantum permanents
is the following identity :
\beqn
QP( (A_{1} \otimes A_{2} ) \rho (A_{3} \otimes A_{4} ) )= \det(A_{1}A_{2}A_{3}A_{4}) QP(\rho)
\eeqn
The author clearly (and sympathetically ) realizes that some readers might object to (or ridicule) the name "quantum permanent". 
The next example, hopefully, will explain possible motivations.
\erem
\xam
Let us present a few cases when Quantum Permanents can be computed "exactly ".
They will also illustrate how universal this new notion is.
\begin{enumerate}
\item
Let $ \rho_{A,B} $ be a product state, i.e. $ \rho_{A,B} = C \otimes D$.
Then $QP(C \otimes D) = N! Det(C) Det(D)$.

\item 
Let $ \rho_{A,B} $ be a pure state, i.e. there exists a matrix  $( R=R(i,j) :  1 \leq i,j \leq N )$ \\
such that $\rho_{A,B}(i_{1},i_{2},j_{1},j_{2}) = R(i_{1},i_{2}) \overline{R(j_{1},j_{2})}$.\\
In this case $QP( \rho_{A,B} ) = N! |Det(R)|^{2} $ . 

\item  
Define blocks of $ \rho_{A,B} $ as  $A_{i,j} = R(i,j) e_{i} e_{i}^{\dagger}$. \\
Then $QP( \rho_{A,B} ) = Per(R) $. 
\end{enumerate}
\exam
The following
propositions provide important upper bounds for quantum permanents of positive semidefinite
matrices.
\pro
Suppose that $\rho_{A,B}$ is a ({\bf BUDM }). Then
\begin{eqnarray}
\max_{\sigma \in S_N} |M(A_{1,\sigma(1)},...,A_{N,\sigma(N)})|  = \nonumber  \\
M(A_{1,1},...,A_{N,N})
\end{eqnarray}
\epro
\prf For $\tau, \sigma \in S_N $ define a matrix \\
$$
B_{\tau, \sigma} = :  A_{\tau(1), \sigma(1)} \otimes A_{\tau(2), \sigma(2)} \otimes... \otimes A_{\tau(N), \sigma(N)}
$$
Since $\rho_{A,B}$ is positive semidefinite hence the block matrix $\{B_{\tau, \sigma} :  \tau, \sigma \in S_N \}$ is
also positive semidefinite. It is well known (\cite{bapat}) and easy to prove that
$$
M(A_1,...,A_N) = tr((A_1  \otimes ...  \otimes A_N) VV^{\dagger})
$$
for some universal $N^{N}$-dimensional vector $V$.\\
It follows that the following $N! \times N!$ matrix $C$
$$
C_{\tau, \sigma} = tr ( B_{\tau, \sigma}VV^{\dagger}) = M(A_{\tau(1), \sigma(1)}, A_{\tau(2), \sigma(2)},...,A_{\tau(N), \sigma(N)})
$$
is also positive semidefinite. Thus 
$$
|C_{\tau, \sigma}| \leq (C_{\tau,\tau}C_{\sigma, \sigma})^{\frac{1}{2}} = M(A_{1,1},...,A_{N,N})
$$
\eprf
\cor
If $\rho_{A,B}$ is ({\bf BUDM })  then
\beqn
QP(\rho_{A,B}) \leq N! M(A_{1,1},...,A_{N,N}) \leq N! Det(\rho_{A}).
\eeqn
The permanental part of Example(3.3) shows that  $N!$ is the
exact constant in both parts of (20),
i.e. if blocks $A_{i,j} =  e_{i}e_{j}^{\dagger}, 1 \leq i,j \leq N$ \\
then $QP(\rho_{A,B}) = N! $ and $M(A_{1,1},...,A_{N,N}) = Det(\rho_{A}) = 1$.
\ecor
The next proposition follows from Hadamard's inequality : \\
if $X \succ 0 $ is $N \times N$ matrix then 
$Det(X) \leq  \prod_{i=1}^{N} X(i,i) $.
\pro 
If $X \succ 0$ then the following inequality holds :
\begin{eqnarray}
Det( \sum_{i=1}^{K}  x_{i}y_{i}^{\dagger} X y_{i}x_{i}^{\dagger} ) \geq  \nonumber \\
 Det(X) MP_{(X,Y) }.
\end{eqnarray}
\epro
\cor
Suppose that a separable ({\bf BUDM })  $\rho_{A,B}$  is Choi's representation of the completely positive operator $T$. \\
Then for all $X \succ 0$ the following inequality holds :
\beqn
Det(T(X)) \geq QP(\rho_{A,B}) Det(X)
\eeqn
Since $\rho_{A} = T(I)$, hence   $QP(\rho_{A,B}) \leq Det(\rho_{A})$  in 
the separable case. \\
(Notice that Corollary 3.5  provides an example of an entangled \\
 ({\bf BUDM }) which does not satisfy (22) .)
\ecor

Finally, in the next theorem we connect  quantum permanents with $G$-norms of determinantal polynomials.
\thm
\begin{enumerate}
\item Consider an arbitrary polynomial
$$
P(x_1,x_2,..., x_k) = \sum_{(r_1,...,r_k) \in F} a_{r_1,...,r_k} x_{1}^{r_1} x_{2}^{r_2}...x_{k}^{r_k},   |F| < \infty
$$
where $F$ is some finite set of vectors  with nonnegative integer components and define its $G$ -norm as follows
$$
 \|P\|_{G}^{2} = :  \sum_{(r_1,...,r_k) \in F} |a_{r_1,...,r_k}|^{2} r_{1}! r_{2}!...r_{k}! 
$$
Then the following identity holds :
\beqn
 \|P\|_{G}^{2} = E _{\xi_{1},...,\xi_{k}}(|P(\xi_{1},...,\xi_{k})|^{2}),
\eeqn
where $(\xi_{1},...,\xi_{k})$ are independent identically distributed zero mean gaussian complex random variables and
the covariance matrix of $\xi_{1}$, viewed as a 2-dimensional real vector, is equal to $\frac{1}{2} I$.
\item
Consider a $k$-tuple of  $N \times N$ complex matrices $ {\bf A} = (A_1,...,A_k) $ and the corresponding determinantal polynomial
$P_{{\bf A}}(x_1,...,x_k) = :  \det( \sum_{1 \leq i \leq k}  x_{i}A_{i} )$. Then the following identity holds
\beqn
\|P_{{\bf A}}\|_{G}^{2}  = QP(\rho_{{\bf A}}  )
\eeqn
\end{enumerate} 
\ethm
\prf The proof is in Appendix 1.
\eprf

\rem
Theorem 3.8, more precisely the combinations of its two parts,can be viewed as a generalization of the famous Wick formula \cite{zvon} \\
It seems reasonable to predict that formula(24) might be of use
in the combinatorics described in \cite{zvon}. \\
It is well known (see, for instance, \cite{Bar1} )  that the mixed discriminant $M(A_1,...,A_{N})$ can be 
evaluated by computing $2^{N}$ determinants. Therefore there the quantum permanent $QP(\rho)$ can be
evaluated by computing $N!2^{N}$ determinants. Now, formula (24) suggests the following algorithm to
compute \\
$\|\det( \sum_{1 \leq i \leq k}  x_{i}A_{i} ) \|_{G}^{2}$ :  \\
first, construct the associated bipartite density
matrix $\rho_{{\bf A}}$, which will take $O(N^{4}k)$ additions and multiplications ; secondly,  compute $QP(\rho_{{\bf A}}  )$.\\
Total cost is $Cost(N) = O(N!2^{N}N^{3})$. On the other hand, just
the number of monomials in $\det( \sum_{1 \leq i \leq k}  x_{i}A_{i} )$
is equal to $|I_{k,N}| = \frac{(N+k-1)!}{N! (k-1)!}$. If $k-1 = aN^{2}$ then  $|I_{k,N}| \geq \frac{a^{N}N^{2N}}{N!}$. 
Thus, 
$$
\frac{|I_{k,N}|}{Cost(N)} \geq \frac{a^{N}N^{2N}}{O(N!^{2}2^{N}N^{3})} \geq \approx \frac{(ae^{2})^{N}}{2^{N}N^{3}}
$$
We conclude that if $a > \frac{2}{e^{2}}$ our approach is exponentially
faster than the "naive" one,  i.e. than evaluating $\det( \sum_{1
\leq i \leq k} x_{i}A_{i} )$ at all vectors $(x_1,...,x_k) \in
I_{k,N}$. \\ Our approach provides an $O(N!2^{N}N^{3})$ - step
deterministic algorithm to solve a general case of Edmonds' Problem
. \\ 
\erem

\section { Operator Sinkhorn's iterative scaling }

Recall that for a square matrix $A = \{a_{ij}: 1 \leq i,j \leq N\}$ row
scaling is
defined as
$$
R(A) = \{ \frac{a_{ij}}{\sum_j a_{ij}} \} , $$
column scaling as $C(A) = \{ \frac{a_{ij}}{\sum_i a_{ij}} \}$
assuming that all
denominators are nonzero.

The iterative process $...CRCR(A)$ is called {\em Sinkhorn's iterative scaling} (SI).
There are two main, well known,  properties of this iterative process, which we will generalize to positive
operators.
\pro
\begin{enumerate}
\item Suppose that $A = \{a_{i,j} \geq 0: 1 \leq i, \ j \leq N\}$.  Then (SI)
converges iff $A$ is matching, i.e., there exists a permutation $\pi$ such
that $a_{i,\pi (i)} > 0 \ (1 \leq i \leq N)$.
\item If $A$ is indecomposable, i.e., $A$ has a doubly-stochastic pattern
and is fully indecomposable in the usual sense, then (SI) converges exponentially
fast. Also in this case  there exist unique positive diagonal matrices $D_1, D_2, \det(D_2) = 1$ 
such that the matrix $D_1^{-1}AD_2^{-1}$ is doubly stochastic.
\end{enumerate}
\epro

\dfn [Operator scaling ]
Consider a positive linear operator $T : M(N) \rightarrow M(N)$. Define a new positive operator, Operator scaling, $S_{C_{1},C_{2}}(T) $ as :
\beqn
S_{C_{1},C_{2}}(T)(X) =: C_{1}T(C_{2}^{\dagger}XC_{2})C_{1}^{\dagger}
\eeqn
Assuming that both $T(I)$ and $T^{*}(I)$ are nonsingular we define analogs of row and column scalings :

\beqn
R(T) = S_{T(I)^{-\frac{1}{2}},I}(T), C(T)= S_{I,T^{*}(I)^{-\frac{1}{2}}}(T)
\eeqn
Operator Sinkhorn's iterative scaling (OSI) is the iterative process $...CRCR(T)$
\edfn
\rem Using Choi's representation of the operator $T$ as in Definition(2.1), we can define analogs of operator scaling (which are exactly
so called local transformations in Quantum Information Theory) and (OSI)  in terms of ({\bf BUDM }) : \\
\begin{eqnarray}
S_{C_{1},C_{2}}( \rho_{A,B}) =  C_{1} \otimes C_{2}( \rho_{A,B} ) C_{1} ^{\dagger} \otimes C_{2}^{\dagger}; \nonumber \\
R( \rho_{A,B}) = \rho_{A}^{-\frac{1}{2}} \otimes I  (\rho_{A,B} ) \rho_{A}^{-\frac{1}{2}} \otimes I, \nonumber \\
C( \rho_{A,B}) =      I   \otimes   \rho_{B}^{-\frac{1}{2}} ( \rho_{A,B} ) I   \otimes   \rho_{B}^{-\frac{1}{2}}.
\end{eqnarray}
The standard ("classical")  {\em Sinkhorn's iterative scaling} is a particular case of Operator Sinkhorn's iterative scaling (OSI) 
when the initial Choi's representation of the operator $T$ is a diagonal ({\bf BUDM }) .
\erem
Let us introduce a class of locally scalable functionals ({\bf LSF }) defined on a set of positive linear operators,
i.e. functionals satisfying the following identity :
\beqn
\varphi(S_{C_{1},C_{2}}(T)) = Det(C_{1}C_{1}^{\dagger}) Det(C_{2}C_{2}^{\dagger}) \varphi(T)
\eeqn
We will call an ({\bf LSF }) bounded if there exists a function $f$ such that $|\varphi(T)| \leq f(tr(T(I))$.
It is clear that bounded ({\bf LSF }) are natural "potentials" for analyzing  (OSI). Indeed, 
Let $T_{n}, T_{0} = T$ be a trajectory of (OSI). $T$ is a positive linear operator. Then $T_{i}(I)=I$ for odd $i$
and $T_{2i}(I)^{*}=I , i \geq 1$. Thus if $\varphi(.)$ is ({\bf LSF })  then
\begin{eqnarray}
\varphi(T_{i+1}) = a(i) \varphi(T_{i}),  a(i) = Det(T_{i}^{*}(I))^{-1} \mbox{ if  } i \mbox{  is odd }, \nonumber \\
a(i) = Det(T_{i}(I))^{-1} \mbox{ if } i > 0 \mbox{  is even}.
\end{eqnarray}
As $tr(T_{i}(I)) = tr(T_{i}^{*}(I)) = N, i > 0$, thus by the arithmetic/geometric means inequality  we have that
$|\varphi(T_{i+1})|  \geq  |\varphi(T_{i})| $   and if $\varphi(.)$ is bounded  and $|\varphi(T)|  \neq 0$
then $DS(T_{n})$ converges to zero. \\

To prove a generalization of Statement 1 in Prop.(4.1) we need to "invent" a bounded ({\bf LSF })   $\varphi(.)$
such that  $\varphi(T) \neq 0$ iff the operator $T$ is rank non-decreasing. We call such functionals ``responsible for matching''.
It follows from (10) and (20) that $QP(CH(T))$ is a bounded ({\bf LSF }). Thus if $QP(CH(T)) \neq 0$ then $DS(T_{n})$ converges to zero 
and, by Prop. (2.4),  $T$ is rank non-decreasing. On the other hand, $QP(CH(Sk_{3})) = 0$  and $Sk_{3}$ is
rank non-decreasing (even indecomposable ).  This is another "strangeness" of entangled operators.
We wonder if it is possible to have a 
"nice",  say polynomial with integer coefficients, responsible for matching ({\bf LSF }) ?
We introduce below a responsible for matching  bounded ({\bf LSF }) which  is continuous  but non-differentiable.
\dfn
For a positive operator $T: M(N) \rightarrow M(N)$, we define its capacity as  
$
Cap(T) = \inf \{ Det(T(X)) : X \succ 0, Det(X) = 1\}. 
$
\edfn
It is easy to see that $Cap(T)$ is ({\bf LSF }).  \\
Since $Cap(T) \leq Det(T(I) ) \leq (\frac{tr(T(I) )}{N})^{N}$, \\
hence $Cap(T)$ is a bounded ({\bf LSF }). \\
\begin{lemma} A positive operator $T: M(N) \rightarrow M(N)$ is rank non-decreasing iff
$Cap(T) > 0$.
\end{lemma}
\prf
Let us fix an orthonormal basis (unitary matrix) $ U =\{u_{1},...,u_{N}\} $ in $C^{N}$ 
and associate with a
positive operator $T$ the following positive operator :
\begin{equation}
T_{U}(X) = :\sum_{1 \leq i \leq N} T(u_{i}u_{i}^{\dagger}) tr(Xu_{i}u_{i}^{\dagger}).
\end{equation}
(In physics terms, $T_{U}$ represents
decoherence with respect to the basis $U$, i.e.
in this basis applying $T_{U}$ to matrix $X$ is the same as applying $T$ to the
diagonal restriction of $X$. )\\
It is easy to see that a positive operator $T$ is rank non-decreasing iff the
operators $T_{U}$ are rank non-decreasing for all unitary $U$. \\
And for fixed $U$ all properties of $T_{U}$ are defined by the following $N$-tuple
of $N \times N$ positive semidefinite matrices :
\begin{equation}
{\bf A}_{T,U} = : (T(u_{1}u_{1}^{\dagger}),..., T(u_{N}u_{N}^{\dagger}).
\end{equation}
Importantly for us,  $T_{U}$ is rank non-decreasing iff the mixed discriminant
$M(T(u_{1}u_{1}^{\dagger}),..., T(u_{N}u_{N}^{\dagger}) ) > 0 $. \\
Define the capacity of ${\bf A}_{T,U}$,  
\begin{eqnarray*}
& Cap({\bf A}_{T,U}) = : \\
& \inf \{Det(\sum_{1 \leq i \leq N} T(u_{i}u_{i}^{\dagger}) \gamma_{i}) : \gamma_{i} > 0, \prod_{1 \leq i \leq N}\gamma_{i} = 1 \}.
\end{eqnarray*}
It is clear from the definitions that    $Cap(T) $ is equal to infimum of $Cap({\bf A}_{T,U})$ over all unitary $U$. \\
One of the main results of \cite{GS} states that
\begin{eqnarray}
M({\bf A}_{T,U}) &=: &M(T(u_{1}u_{1}^{\dagger}),..., T(u_{N}u_{N}^{\dagger}) ) \leq Cap({\bf A}_{T,U}) \leq \nonumber \\
& & \leq  \frac{N^{N}}{N!} M(T(u_{1}u_{1}^{\dagger}),..., T(u_{N}u_{N}^{\dagger}) ).
\end{eqnarray}
As the mixed discriminant is a continuous (analytic) functional and the group $SU(N)$ of unitary matrices is compact,
we get the next inequality:
\beqn
\min_{U \in SU(N) } M({\bf A}_{T,U}) \leq Cap(T) \leq \frac{N^{N}}{N!}\min_{U \in SU(N) }M({\bf A}_{T,U})
\eeqn
The last inequality proves that $Cap(T) > 0$ iff positive operator$T$ is rank non-decreasing.
\eprf

So, the capacity is a bounded ({\bf LSF }) responsible for matching, which proves the next theorem :
\thm 
\begin{enumerate}
\item
Let $T_{n}, T_{0} = T$ be a trajectory of (OSI), where 
$T$ is a positive linear operator. Then $DS(T_{n})$ converges to zero iff
$T$ is rank non-decreasing.
\item
A positive linear operator $T : M(N) \rightarrow M(N)$ is rank non-decreasing iff for all $\epsilon > 0$ there exists an $\epsilon$-doubly stochastic
operator scaling of $T$.
\item
A positive linear operator $T$ is rank non-decreasing iff there exists $\frac{1}{N}$-doubly stochastic
operator scaling of $T$.
\end{enumerate}
\ethm
The next theorem generalizes second part of Prop. (4.1) and  is proved on almost the same lines as Lemmas 24,25,26,27 in \cite{GS}.
\thm
\begin{enumerate}
\item  There exist nonsingular matrices $C_{1},C_{2}$    such that $S_{C_{1},C_{2}}(T)$ is doubly stochastic
iff the infimum in Definition 4.4  is attained.  \\
Moreover,  if $Cap(T)  = Det(T(C)) $ where $ C \succ 0,  Det(C)=1 $   \\
then $S_{T(C)^{\frac{-1}{2}}, C^{\frac{1}{2}}} (T)$  is doubly stochastic. \\
Positive operator $T$ is indecomposable iff the infimum in Definition 4.4 is attained and unique.
\item  A doubly stochastic operator $T$ is indecomposable iff  \\
 $ tr(T(X))^{2} \leq a \ tr(X)^{2}$ for some $0 \leq a < 1$
and all traceless hermitian matrices $X$.
\item  If a positive operator $T$ is indecomposable then $DS(T_{n})$ converges to zero with the exponential rate,
i.e. $DS(T_{n}) \leq K a^{n}$ for some $K$ and $0 \leq a < 1$.
\end{enumerate}
 \ethm
\rem
Consider an $N \times N$ matrix $A$ with nonnegative entries. Similarly to (30), define its
capacity as follows :
$$
Cap(A) = \inf \{ \prod_{1 \leq i \leq N} (Ax)_{i}  : x_{i} > 0, 1 \leq i \leq N ;  \prod_{1 \leq i \leq N} x_{i}= 1\} 
$$
Recall that the KLD-divergence between two matrices is defined as \\
$$
KLD(A||B) = \sum_{1 \leq i,j \leq N} B(i,j) \log (\frac{B(i,j)}{A(i,j)})
$$
It is easy to prove (see, for instance, \cite{GY} ) that  
$$
-\log (Cap(A) ) = \inf \{KLD(A||B) :  B \in D_{N} \},
$$
where $D_{N}$ is the
convex compact set of $N \times N$ doubly stochastic matrices. \\
Of course, there is a quantum analog of KLD-divergence, the
so called von Neumann divergence. It is not clear
whether there exists a similar "quantum" characterization of the capacity  of completely positive operators. \\
The inequality (20) can be strengthen to the following one :
$$
QP(CH(T)) \leq N! Cap(T) 
$$
And $N!$ is also an exact constant in this inequality above.
\erem
\section{Polynomial time deterministic algorithm for the Edmonds Problem}
Let us consider the following three properties of ({\bf BUDM })  $\rho_{A,B}$.
( We will view this $\rho_{A,B} $  as Choi's representation of a 
completely positive 
operator $T$, i.e.      $\rho_{A,B} =   CH(T)$. )
\begin{description}
\item[P1] 
$Im(\rho_{A,B})$  contains a nonsingular matrix. 
\item [P2]
The Quantum permanent $QP(\rho_{A,B}) > 0$. 
\item [P3] 
Operator $T$ is rank non-decreasing.
\end{description}
Part 2 of theorem (3.8) proves that $P1 \Leftrightarrow P2 $ and
Example(2.8) illustrated that that the implication $P2 \Rightarrow P3$
is strict.  It is not clear whether either $P1 $ or $P3$ can be
checked in deterministic polynomial time. \\ Next, we will describe
and analyze Polynomial time deterministic algorithm to check whether
$P3$ holds provided that it is promised that $Im(\rho_{A,B})$ , viewed
as a linear subspace of $M(N)$, has the {\bf Edmonds-Rado Property }.
Or, in other words, that it is promised that $P1 \Leftrightarrow
P3$. \\ In terms of Operator Sinkhorn's iterative scaling (OSI) we
need to check if there exists $n$ such that $DS(T_{n}) \leq
\frac{1}{N} $. If $ L = : \min\{n: DS(T_{n}) \leq \frac{1}{N} \} $ is
bounded by a polynomial in $N$ and number of bits of $\rho_{A,B}$ then
we have a Polynomial time Deterministic algorithm to solve Edmonds'
problem provided that it is promised that $P1 \Leftrightarrow P3$.
Algorithms of this kind for "classical" matching problem appeared
independently in \cite{lsw} and \cite{GY}.  In the "classical" case
they are just another, conceptually simple, but far from optimal,
poly-time algorithms to check whether a perfect matching exists.  But
in this general Edmonds Problem setting, our, Operator Sinkhorn's
iterative scaling based approach seems perhaps to be the only
possibility. \\ Assume, without loss of generality, that all entries
of $\rho_{A,B}$ are integer numbers and their maximum magnitude is
$\bf{M}$. Then $Det(\rho_{A}) \leq (\bf{M}N)^{N}$ by Hadamard's inequality.  If
$QP(\rho_{A,B}) > 0$ then necessary $QP(\rho_{A,B}) \geq 1$ for it is
an integer number.  Thus
$$
QP(CH(T_{1})) = \frac{QP(CH(T))}{Det(\rho_{A})} \geq (\bf{M}N)^{-N}.
$$

Each $nth$ iteration ($n \leq L$ ) after the first one will multiply the Quantum permanent by $Det(X)^{-1}$, 
where $X \succ 0, tr(X)=N$ and $tr((X-I)^{2}) > \frac{1}{N} $. Using results from \cite{lsw},
$Det(X)^{-1} \geq (1 -  \frac{1}{3N})^{-1} = : \delta $. Putting all this together,
we get the following upper bound on $L$, the number of steps in (OSI) to reach the "boundary"
$DS(T_{n}) \leq \frac{1}{N}$  :
\beqn
\delta^{L}  \leq \frac{QP(CH(T_{L}))}{(\bf{M}N)^{-N}}
\eeqn
It follows from (20)  that  $QP(CH(T_{L})) \leq N!$ \\

Taking logarithms we get that 
\beqn
L \leq \approx  3N(N \ln(N) +N(\ln(N) + \ln(\bf{M})) ;
\eeqn

Thus $L$ is polynomial in the dimension  $N$ and the number of bits $\log(\bf{M})$. \\
To finish our analysis, we need to evaluate the complexity of each step of (OSI). \\
Recall that $T_{n}(X) = L_{n}(T(R_{n}^{\dagger}XR_{n}))L_{n}^{\dagger}$ \\
for some nonsingular matrices $ L_{n}$ and $R_{n}$, \\
$T_{n}(I) =  L_{n}(T(R_{n}^{\dagger}R_{n}))L_{n}^{\dagger}$ and 
$T_{n}^{*}(I) =  R_{n}(T^{*}(L_{n}^{\dagger}L_{n}))R_{n}^{\dagger}$. \\
To evaluate $DS(T_{n})$ we need to compute $tr((T_{n}^{*}(I)-I)^{2})$ for odd \\
$n$  and $tr((T_{n}(I)-I)^{2})$ for even $n$. \\
Define $P_{n}=L_{n}^{\dagger}L_{n},  Q_{n} = R_{n}^{\dagger}R_{n} $.
It is easy to see that the matrix $T_{n}(I)$ is similar to $P_{n}T(Q_{n})$,
and $T_{n}^{*}(I)$ is similar to $Q_{n}T^{*}(P_{n})$.  \\
As traces of similar matrices are equal, to evaluate $DS(T_{n})$
it is sufficient to compute matrices $P_{n}, Q_{n} $. \\
But, $ P_{n+1} = (T(Q_{n}))^{-1} $   and  $ Q_{n+1} = (T^{*}(P_{n}))^{-1} $. \\
And this leads to  standard rational matrix operations with $O(N^{3})$ per 
iteration in (OSI). \\
Notice that our original definition of (OSI) requires computation of an operator square root.
It can be replaced by the Cholesky factorization, which still requires computing scalar
square roots. But our final algorithm is rational!
\rem
To ensure that all the matrices we need to invert along the algorithm (OSI) are nonsingular indeed  it is sufficient 
that both $T(I) \succ 0$ and $T^{*}(I) \succ 0$ (strictly positive definite) . 
It is easy to see  that if positive operator $T$ is rank non-decreasing then its dual $T^{*}$ is also
rank non-decreasing . Thus if positive operator $T$ is rank non-decreasing then necessarily 
$T(X) \succ 0$ and $T^{*}(X) \succ 0$ for all $X \succ 0 $ .
\erem

\section{Weak Membership Problem for the convex compact set of normalized
bipartite separable density matrices is NP-HARD }
One of the main research activities in Quantum Information Theory is a search for "operational"
criterion for the separability.  We will show in this section that, in a sense defined below, the problem is
NP-HARD even for bipartite normalized density matrices provided that each part is large (each "particle"
has large number of levels). First, we need to recall some basic notions from computational convex geometry.
\subsection{Algorithmic aspects of convex sets }
We will follow \cite{gr:lo:sc}.
\dfn
A proper (i.e. with nonempty interior) convex set $K \subset R^{n}$ is called well-bounded $a$-centered
if there exist a rational vector $a \in K$  and positive  (rational) numbers $r,R$ such that
$B(a,r) \subset K$  and $K \subset B(a,R)$    (here $B(a,r) = \{x : \|x-a\| \leq r \}$ and $\|. \|$ is a standard euclidean
norm in $R^{n}$ ). The encoding length of such a convex set $K$  is
$$
<K> = n + <r> + <R> + <a> ,
$$
where  $<r>, <R>, <a>$  are the number of bits of corresponding rational numbers and rational vector.\\
Following \cite{gr:lo:sc} we define $S(K,\delta)$ as a union of all $\delta$-balls with centers belonging to $K$ ;
and $S(K,-\delta) = \{ x \in K :  B(x,\delta) \subset K \} $. \\
\edfn
\dfn 
The Weak Membership Problem ($WMEM(K,y,\delta)$) is defined as follows : \\
Given a rational vector $y \in R^{n}$ and a rational number $\delta > 0$ either \\
(i) assert that $y \in S(K,\delta)$, or \\
(ii) assert that $y \not\in S(K,-\delta)$. \\

The Weak Validity Problem ($WVAL(K,c, \gamma, \delta ) $) is defined as follows : \\
Given a rational vector $c \in R^{n}$, rational number $\gamma$ and a rational number $\delta > 0$ either \\
(i) assert that $ <c,x> = : c^{T}x \leq \gamma + \delta $ for all $x \in S(K,-\delta)$, or \\
(ii) assert that $ c^{T}x \geq \gamma - \delta $  for some $x \in S(K,\delta)$. \\
\edfn
\rem
Define $M(K,c) = : \max_{x \in K} <c,x>$ . It is easy to see that
\begin{eqnarray*}
& M(K,c) \geq M(S(K,-\delta),c) \geq M(K,c) - \|c\| \delta \frac{R}{r} ;  \\
 & M(K,c) \leq M(S(K,\delta),c) \leq M(K,c) + \|c\| \delta
\end{eqnarray*}
\erem
Recall that the seminal Yudin-Nemirovski theorem (\cite{YN}, \cite{gr:lo:sc}) implies that if there exists a deterministic
algorithm solving $WMEM(K,y,\delta)$  in $Poly( <K> + <y> + <\delta>)$  steps  then there exists a deterministic
algorithm solving  $WVAL(K,c, \gamma, \delta ) $ in $Poly( <K> + <c> +  <\delta>  + <\gamma>)$ steps. \\
Let us denote as $SEP(M,N)$ a compact convex set of separable density matrices $\rho_{A,B} : C^{M} \otimes C^{N} \rightarrow C^{M} \otimes C^{N} $,
$tr(\rho_{A,B})=1$,  $M \geq N$.  Recall that 
\begin{eqnarray*}
& SEP(M,N) =  \\
& CO(\{xx^{\dagger} \otimes yy^{\dagger} :  x \in C^{M},y \in C^{N}; \|x\| =  \|y\| = 1 \}),
\end{eqnarray*}
where $CO(X)$ stands for the convex hull generated by a set $X$.\\
Our goal is to prove that the 
Weak Membership Problem for $SEP(M,N)$ is NP-HARD.
As we are going to use the Yudin--Nemirovski theorem, it is sufficient to prove that $WVAL(SEP(M,N),c, \gamma, \delta ) $ is NP-HARD with respect
to the complexity measure $(M + <c> +  <\delta>  + <\gamma>)$
and to show that $<SEP(M,N)>$ is polynomial in $M$.
\subsection {Geometry of $SEP(M,N)$  }
First,  $SEP(M,N)$ can be viewed as a compact convex subset of the hyperplane in $R^{D}, D=: N^{2}M^{2}$. The standard euclidean 
norm in $R^{N^{2}M^{2}}$ corresponds
to the Frobenius norm for density matrices, i.e. $\|\rho\|_{F} = tr (\rho \rho^{\dagger})$.  The matrix $\frac{1}{NM} I  \in SEP(M,N)$
and $\|\frac{1}{NM} I -  xx^{\dagger} \otimes yy^{\dagger}\|_{F}  =  \sqrt{\frac{D-1}{D}} < 1 $ for all norm one
vectors $x,y$. Thus $SEP(M,N)$ is covered by the ball $B(\frac{1}{NM} I, \sqrt{\frac{D-1}{D}})$.  \\
The following result was recently proved in \cite{GB2002}.
 \thm
Let $\Delta$ be a block hermitian matrix as in (5). If $tr(  \Delta ) = 0$ and 
 $\|\Delta\|_{F} \leq \sqrt{\frac{1}{D(D-1)}}$ then the the block matrix $\frac{1}{D} I + \Delta$ is
separable.
\ethm
Summarizing, we get that  for $D=MN$ 
$$
B(\frac{1}{D} I, \sqrt{\frac{1}{D(D-1)}}) \subset SEP(M,N) \subset B(\frac{1}{D} I,\sqrt{\frac{D-1}{D}}),
$$
(balls are restricted to the corresponding hyperplane ) 
and conclude that $<SEP(M,N)> \leq  Poly(MN) $.  
It is left to prove that $WVAL(SEP(M,N),c, \gamma, \delta ) $ is NP-HARD with
respect
to the complexity measure $(MN+ <c> +  <\delta>  + <\gamma>)$.
\subsection{ Proof of Hardness }
Let us consider the following hermitian block matrix :
\begin{equation} 
C = \left( \begin{array}{cccc}
		  0 & A_{1} & \dots & A_{M-1}\\
		  A_{1} & 0 & \dots & 0\\
		  \dots &\dots & \dots & \dots \\
		  A_{M-1} & 0 & \dots & 0\end{array} \right),
\end{equation}
i.e. its $(i,j)$ blocks are zero if either $i \neq 1$ or $j \neq 1$ and $(1,1)$ block is also zero ;   $A_{1},..., A_{M-1}$
are real  symmetric $N \times N$ matrices.
\pro
\begin{eqnarray*}
& \max_{\rho \in SEP(M,N)} (tr (C \rho))^{2} = \\
& \max_{y \in R^{N}, \|y\| = 1}    \sum_{1 \leq i \leq M-1} (y^{T}A_{i}y)^{2}.
\end{eqnarray*}
\epro
\prf 
First, by linearity and the fact that the set of extreme points 
\begin{eqnarray*}
& Ext(SEP(M,N)) = \\
& \{ xx^{\dagger} \otimes yy^{\dagger} : x \in C^{M},y \in C^{N} ; \|x\| =  \|y\| = 1 \}
\end{eqnarray*}
we get that 
\begin{eqnarray*}
& \max_{\rho \in SEP(M,N)} tr (C \rho)  = \\
 & \max_{ xx^{\dagger} \otimes yy^{\dagger} :  x \in C^{M},y \in C^{N}; \|x\| =  \|y\| = 1 }  tr(C (xx^{\dagger} \otimes yy^{\dagger})).
\end{eqnarray*}
But $tr(C (yy^{\dagger} \otimes xx^{\dagger})) = tr (A(y) xx^{\dagger})$, where real symmetric $M \times M$ matrix $A(y)$ is defined as follows :
$$
A(y) = \left( \begin{array}{cccc}
		  0 & a_{1} & \dots & a_{M-1}\\
		  a_{1} & 0 & \dots & 0\\
		  \dots &\dots & \dots & \dots \\
		  a_{M-1} & 0 & \dots & 0\end{array} \right)  ;    
$$		  
$ a_{i} = tr (A_{i}  yy^{\dagger}), 1 \leq i \leq M-1. $ \\
Thus 
\begin{eqnarray*}		  
& \max_{\rho \in SEP(M,N)} tr (C \rho) = \\
& \max_{ yy^{\dagger} \otimes xx^{\dagger} : x \in C^{M},y \in C^{N} ; \|x\| =  \|y\| = 1 } tr (C(xx^{\dagger} \otimes yy^{\dagger}) )   = \\
& \max_{\|y\| = 1 }  \lambda_{max} A(y).
\end{eqnarray*}
(Above $\lambda_{max} A(y)$ is the maximum eigenvalue of $A(y)$) \\
It is easy to see $A(y)$ has only two real non-zero eigenvalues  \\
$( d, -d )$, where $d= \sum_{1 \leq i \leq M-1} (tr(A_{i}  yy^{\dagger}))^{2} $.\\
As $A_{i}, 1 \leq i \leq N-1 $ are real symmetric matrices we finally get that 
$$
\max_{\rho \in SEP(M,N)} ( tr (C \rho))^{2} = \max_{y \in R^{N}, \|x\| = 1} \sum_{1 \leq i \leq N-1} (y^{T}A_{i}y)^{2}.
$$
\eprf

Proposition(6.5) and Remark(6.3) suggest that in order  to prove NP-HARDness of \\
 $WVAL(SEP(M,N),c, \gamma, \delta ) $  with respect
to the complexity measure $M + <c> +  <\delta>  + <\gamma>$ it is sufficient to prove that the following problem is NP-HARD : \\
\dfn (RSDF problem)  Given  $k $  $l \times l$ real rational symmetric matrices $( A_{i}, 1 \leq i \leq l )$  and rational numbers $(\gamma, \delta)$ to check
whether
$$
\gamma + \delta \geq max_{x \in R^{l}, \|x\| = 1}f(x)  \geq \gamma - \delta,  f(x) = :   \sum_{1 \leq i \leq l} (x^{T}A_{i}x)^{2}.
$$
respect to the complexity measure \\
$(l  k + \sum_{1 \leq i \leq l}<A_{i}>  + <\delta>  + <\gamma>)$.
\edfn
It was shown in \cite{BN}, by a reduction from KNAPSACK,  that the
RSDF problem is NP-HARD provided\\
 $k \geq \frac{l(l-1)}{2} + 1$. \\
We summarize all this in the following theorem
\thm
The Weak Membership Problem for $SEP(M,N)$ is NP-HARD if   $N \leq M \leq \frac{N(N-1)}{2} + 2$.
\ethm 
\rem
It is easy exercise to prove that ({\bf BUDM }) $\rho_{A,B}$ written in block form (3) is real separable iff
it is separable and all the blocks in (3) are real symmetric matrices. It follows that, with obvious modifications,
Theorem 6.7  is valid for real separability too. \\
The construction (37) was inspired by Arkadi Nemirovski's proof of 
the NP-HARDness of checking the positivity
of a given operator \cite{nem}.
\erem
\section{Concluding Remarks}
Many ideas of this paper were suggested by \cite{GS}.  The world of mathematical
interconnections is very unpredictable (and thus is so exciting). The main technical result in a very recent breakthrough in
Communication Complexity \cite{fos} is a rediscovery of particular, rank one, case of a general, matrix tuples scaling,
 result proved in \cite{GS} with much simpler proof than in \cite{fos}.
Perhaps this paper will produce something new in Quantum Communication Complexity. \\
We still don't know whether there is a deterministic poly-time algorithm to check whether a given
completely positive operator is rank non-decreasing. And this question is related to lower bounds
on $Cap(T)$ provided that Choi's representation $CH(T) $ is an integer semidefinite matrix. \\
Theorem(6.7)  together with other
results from our paper gives a new, classical complexity based, insight on the nature
of quantum entanglement and, in a sense, closes a long line of research in
Quantum Information Theory.   \\
Also, this paper suggests a new way to look at "the worst entangled" bipartite density matrices (or completely
positive operators). For instance, the operator $Sk_{3}$ from Example (2.8) seems to be "the  worst entangled" and
it is not surprising that it appears in many counterexamples.\\
We hope that the constructions introduced in this paper, especially the
Quantum Permanent,
will have a promising future. \\
We think, that in general, mixed discriminants and mixed volumes \cite{Alexandrov} should be studied and used
more enthusiastically in the Quantum context.
After all, they are noncommutative generalizations of the permanent.... \\
The $G$-norm defined in (2) appears in this paper mainly because of formula (24). 
It is called by some authors ( \cite{zelin} ) Bombieri's norm (see also, \cite{BBEM}, \cite{reznick}, \cite{beau} ). \\
Also, the $G$-norm arises naturally in quantum optics and the study of quantum harmonic
oscillators. \\
This norm satisfies some remarkable properties ( \cite{BBEM}, \cite{reznick}) which, we think, can be used
in quantum/linear optics computing research. \\
Combining formulas (23) and (24) , one gets an unbiased nonnegative valued random estimator for quantum permanents
of bipartite unnormalized density matrices.  
But, as indicated in \cite{laur}, it behaves rather badly for
the entangled bipartite unnormalized density matrices. \\
From the other hand,  there is a hope, pending on a proof of a "third" generation of van der Waerden \\
conjecture  ((\cite{fr}, \cite{fal}, \cite{ego} ),  (\cite{GS}, \cite{gur})) ,
to have even a deterministic polynomial time algorithm
to approximate within a simply exponential factor quantum permanents of separable unnormalized bipartite density matrices 
(more details on this matter can be found in \cite{quant}).
It is my great pleasure to thank my LANL  colleagues Manny Knill and Howard Barnum . \\
Many thanks to Marek Karpinski and Alex Samorodnitsky for their comments on this paper . \\
Finally, I would like to thank Arkadi Nemirovski for many enlightening discussions.
\appendix
\section{Proof of Theorem (3.8)  and a permanental corollary }
The main goal of this Appendix is a "direct proof" of formula (24) . A much shorter
probabilistic proof is presented in Appendix {\bf C }. \\
\prf [ Proof of formula (23) ] \\
It is sufficent to prove that for any monomial 
\beqn
 \frac{1}{\pi^{k}} \int ...\int  |z_{1}^{r_1}...z_{k}^{r_k}|^{2} e^{-(x_{1}^{2} + y_{1}^{2})}...e^{-(x_{k}^{2} + y_{k}^{2})}
dx_{1} d y_{1} ... dx_{k} d y_{k} = r_{1}! r_{2}! ...r_{k}!  (z_{l} = x_{l} + i y_{l} , 1 \leq l \leq k ) .
\eeqn
And that distinct monomials are orthogonal , i.e. 
\beqn
\int ...\int  (z_{1}^{r_1}...z_{k}^{r_k} \overline{z_{1}^{h_1}...z_{k}^{h_k}} )  e^{-(x_{1}^{2} + y_{1}^{2})}...e^{-(x_{k}^{2} + y_{k}^{2}) }
dx_{1} d y_{1} ... dx_{k} d y_{k} = 0   ( r \neq h)
\eeqn
Notice that both $2k$-dimensional integrals (37) and (38) are products of corresponding 2-dimensional integrals .
Thus (37) is reduced to the fact that
$$
  \frac{1}{\pi} \int \int (x_{1}^{2} + y_{1}^{2}) ^{2r_{1}} e^{-(x_{1}^{2} + y_{1}^{2})} dx_{1} d y_{1} = r_{1}! .
$$
Using polar coordinates in a standard way , we get that
$$
 \frac{1}{\pi} \int \int (x_{1}^{2} + y_{1}^{2}) ^{2r_{1}} e^{-(x_{1}^{2} + y_{1}^{2})} dx_{1} d y_{1} =
 \int_{0}^{\infty} R^{r_{1}} e^{-R} dR =  r_{1}! .
$$
Similarly (38) is reduced to
$$
\int \int (x_{1} + i  y_{1}) ^{m} (x_{1}^{2} + y_{1}^{2}) ^{k}e^{-(x_{1}^{2} + y_{1}^{2})} dx_{1} d y_{1} = 0 ,
$$
where $m$ is positive integer  and $k$ is nonnegative integer . \\
But
$$
\int \int (x_{1} + i  y_{1}) ^{m} (x_{1}^{2} + y_{1}^{2}) ^{k}e^{-x_{1}^{2} + y_{1}^{2}} dx_{1} d y_{1} =
\int_{0}^{\infty} R^{2k} e^{-R^{2}} ( \int_{0}^{2\pi} e^{-i m \phi} d\phi ) dR = 0 .
$$
\eprf

\prf [Proof of formula (24) ] \\
First , let us recall how coefficients of $\det( \sum_{1 \leq i \leq k}  x_{i}A_{i} )$ can be expressed in terms of
the corresponding mixed discriminants . Let us associate a vector $r  \in I_{k,N}$ an $N$-tuple of $N \times N$ complex matrices ${\bf B}_{r}$ consisting
of  $r_i$ copies of $A_{i}  (1 \leq i \leq k ) $ .\\ 
Notice that 
$$
{\bf B}_{r} = (B_1,...,B_N) ;  B_i \in \{A_1,...,A_k\} ,  1 \leq i \leq k .
$$
It is well known and easy to check that for this particular determinantal polynomial its coefficients satisfy the
following identities :
\beqn
 a_{r_1,...,r_k} =  \frac{M({\bf B}_{r})}{r_{1}! r_{2}! ...r_{k}! }  ;  (r_1,...,r_k) \in I_{k,N} 
 \eeqn
We already defined mixed discriminants by two equivalent formulas (13) , (14) . The next equivalent definition
is handy for our proof :
\beqn
M(B_1,...B_N) =  \sum_{\sigma \in S_N}  \det ([B_{1}(e_{\sigma(1) } | B_{2}(e_{\sigma(2) }| ...|B_{N}(e_{\sigma(N) }]) .
\eeqn
In the formula( 40) above , $(e_1,...,e_N)$ is a canonical basis in $C^{N}$ , and for a $N \times N$ complex matrix $B$
a column vector $B(e_i)$ is an $i$-th column of $B$ . \\
We will use in this proof three basic elementary facts about mixed discriminants . First is "local additivity" , i.e.
$$
M(A_{1}+B, A_{2},...,A_{N}) = M(A_{1}, A_{2},...,A_{N}) + M(B, A_{2},...,A_{N}) .
$$
Second is permutation invariance , i.e .
$$
M(A_{1}, A_{2},...,A_{N}) = M(A_{\tau(1)}, A_{\tau(2)},...,A_{\tau(N)})  , \tau \in S_{N} .
$$
And the third one is easy formula for the rank one case :
$$
M(x_{1}y_{1}^{T},...,x_{N}y_{N}^{T}) = \det (x_{1}y_{1}^{T}+...+x_{N}y_{N}^{T}) ,
$$
where  $( x_{i} , y_{i} ; 1 \leq i \leq N ) $ are $N$-dimensional complex column-vectors . \\
Recall that blocks of $\rho_{{\bf A}}$  are defined as
$$
 A_{i,j} = \sum _{1 \leq i \leq k} A_{k}e_{i}e_{j}^{\dagger}A_{k}^{\dagger} ,  1 \leq i,j \leq N .
$$
Let us rewrite formula (15) as follows :
\beqn
QP( \rho ) = : \frac{1}{N!} \sum_{\sigma , \tau \in S_N} (-1)^{sign(\sigma)}M(A_{\tau(1),\sigma(1)},...,A_{\tau(N),\sigma(N)}) ;
\eeqn

Using this formula (41) we get the following expression for quantum permanent of bipartite density matrix $\rho_{{\bf A}}$ using "local"
additivity of mixed dicriminant in each matrix component :
$$
QP(\rho_{{\bf A}})  = \frac{1}{N!} \sum_{t_1,...,t_N} \sum_{\tau_1 , \tau_2 \in S_{N} } M(A_{t_1}e_{\tau_{1}(1)} 
e_{\tau_{2}(1)}^{\dagger}A_{t_1}^{\dagger} ,..., A_{t_N}e_{\tau_{1}(N)}e_{\tau_{2}(N)}^{\dagger}A_{t_N}^{\dagger} ) .
$$
Using rank one formula above and formula(40) , we get that 
$$
\sum_{\tau_1 , \tau_2 \in S_{N} } M(A_{t_1}e_{\tau_{1}(1)} 
e_{\tau_{2}(1)}^{\dagger}A_{t_1}^{\dagger} ,..., A_{t_N}e_{\tau_{1}(N)}e_{\tau_{2}(N)}^{\dagger}A_{t_N}^{\dagger}) =
|M(A_{t_1},...,A_{t_N})|^{2} .
$$
The last formula gives the following , intermediate , identity :
\beqn
QP(\rho_{{\bf A}})  = \frac{1}{N!} \sum_{t_1,...,t_N} |M(A_{t_1},...,A_{t_N})|^{2} .
\eeqn
What is left is to "collect" in (42) , using invariance of mixed discriminants respect to permutations ,
all occurances of $M({\bf B}_{r})$ (as defined in (39)) , where $ r =(r_1,...,r_{k})  \in I_{k,N} $ . \\
It is easy to see that this number $N(r_1,...,r_{k}) $ of occurances of $M({\bf B}_{r})$ is equal to the coefficient
of monomial  $x_{1}^{r_1} x_{2}^{r_2}...x_{k}^{r_k}$ in the polynomial  $(x_{1}+...+x_{k})^{N}$ . \\
In other words , $N(r_1,...,r_{k}) = \frac{N!}{r_1! ... r_{k}!}$ , which finally gives that 
$$
QP(\rho_{{\bf A}}) = \sum_{ r \in \in I_{k,N}} \frac{|M({\bf B}_{r})|^{2}}{r_1! ... r_{k}!} .
$$
Using formula(39) for coefficients of determinantal polynomial  $\det( \sum_{1 \leq i \leq k}  x_{i}A_{i} )$ we get
that
$$
\|P_{{\bf A}}\|_{G}^{2} = \sum_{(r_1,...,r_k) \in I_{k,N}} |a_{r_1,...,r_k}|^{2} r_{1}! r_{2}! ...r_{k}!  = QP(\rho_{{\bf A}})
$$
\eprf
Putting Part 1 and Part 2  together we get in the next corollary a formula expressing permanents of positive semidefinite
matrices as squares of $G$-norms of multilinear polynomials . A particular , rank two case , of this formula was
 ( implicitely ) discovered in \cite{reznick} .
\cor
Consider complex positive semidefinite $N \times N$ matrix  $Q = DD^{\dagger}$  , where a "factor" $D$ is  $N \times M$ 
complex matrix .  Define a complex gaussian vector  $z = D \xi$ , where $\xi$ is an $M$-dimensional complex gaussian
vector as in theorem 3.8 . \\
The following formula provides unbiased nonnegative valued random estimator for $Per(Q)$ :
\beqn
Per(Q)  = E_{\xi_1 , ..., \xi_N} (|z_1|^{2}...|z_N|^{2}) .
\eeqn
\ecor

\prf
Consider the following $m$-tuple of complex $N \times N$ matrices :
$$
{\bf Diag } = (Diag_{1} , ..., Diag_{m})  ;  Diag_{j} = Diag (D(1,j) , ...,D(N,j)) , 1 \leq j \leq M .
$$
Then $P_{{\bf Diag}} (x_1,...,x_m) = \prod_{1 \leq i \leq N} ( Dx )_{i} $ , where $( Dx )_{i}$ is $i$th component of vector $Dx$ . \\
\
\\
Thus Part 1 of theorem 3.8  gives that $\|P_{{\bf Diag}}\|_{G}^{2} = E_{z_1 , ..., z_N} (|z_1|^{2}...|z_N|^{2}) . $ \\
It is easy to see that the block representation of bipartite density matrix $\rho_{{\bf Diag}}$   associated with $m$-tuple ${\bf Diag }$
is as follows :
$$
\rho_{{\bf Diag}} = \left( \begin{array}{cccc}
		  A_{1,1} & A_{1,2} & \dots & A_{1,N}\\
		  A_{2,1} & A_{2,2} & \dots & A_{2,N}\\
		  \dots &\dots & \dots & \dots \\
		  A_{N,1} & A_{N,2} & \dots & A_{N,N}\end{array} \right) ,       A_{i,j} = Q(i,j)e_{i}e_{j}^{T} .
$$
Therefore $QP(\rho_{{\bf Diag}}) = Per(Q) . $\\
Now Part 2 of theorem 3.8 gives that 
\beqn
Per(Q) = QP(\rho_{{\bf Diag}}) = \|P_{{\bf Diag}}\|_{G}^{2} = E_{z_1 , ..., z_N} (|z_1|^{2}...|z_N|^{2}) . 
\eeqn
\eprf
\rem Corollary (A.1) together with  a remarkable supermultiplicative inequality for the $G$-norm ( \cite{BBEM}, \cite{reznick})
give a completely new look at many nontrivial permanental inequalities , such as famous Leib's inequality \cite{minc}  etc, and
allow new corellational inequalities for analytic functions of complex gaussian vectors  and new ( "short" ) characterizations
of independence of  analytic functions of complex gaussian vectors . More on this will be described in \cite{gurwick} .
\erem
\section{ Wick formula }
In the next theorem we recall famous Wick formula (see , for instance , \cite{zvon} ) .
\thm
Consider complex $ 2N \times M$ matrix  $A$ and  a real $M$-dimensional gaussian vector $x$  with zero mean
and covariance matrix $E(xx^{T}) = I$ .  Define $(y_1,...,y_{2N})^{T} = Ax$ . Then the following Wick formula holds
\beqn
W(A) = : E( \prod_{1 \leq i \leq 2N} y_{i} ) = Haf(AA^{T}) ,
\eeqn
where hafnian $Haf(B)$ of $2N \times 2N$ matrix $B$ is defined as follows :
\beqn
Haf(B) = \sum_{1 \leq p_{1} < p_{2} < ... p_{N} ;  p_{1} < q_{1} , ..., p_{N} < q_{N} \leq 2N }  \prod_{1 \leq i \leq N} B(p_{i},q_{i}) 
\eeqn
\ethm
Let us show how formula (43) follows from (45) .
\pro
Suppose that complex $N \times M$ matrix $D$ in Theorem 1.4 can be written as $D= C+iB$ .
Consider the following complex $2N \times 2M$ matrix $A$ :
$$
\sqrt{2} A=  \left( \begin{array}{cc}
		  C+iB & iC-B \\
		 C-iB & -B -iC \end{array} \right) .		  
$$

Then  

$$
AA^{T} =  \left( \begin{array}{cc}
		  0 & DD^{\dagger} \\
		 DD^{\dagger}& 0 \end{array} \right) ,
$$

 and  $W(A) = E_{z_1 , ..., z_N} (|z_1|^{2}...|z_N|^{2})  $  , where the expression \\
$E_{z_1 , ..., z_N} (|z_1|^{2}...|z_N|^{2})$ is the same as in Corollary (A.1) . \\
\epro
As it easy to see that $Haf(AA^{T}) = Per(DD^{\dagger})$ , thus ,  using Wick formula (45)  we
reprove formula (43) .

 Summarizing , we can say at this point that formula(43) is essentially a different way to write Wick formula .
 (We thank A.Barvinok for pointing at this observation and reference \cite{zvon}). \\
 From the other hand formula(43) is a direct corrolary of  formula(24) for the case of tuples of diagonal matrices .
 It is easy to see that one can also consider upper triangular matrices . \\
 More generally consider the following group action on tuples of square complex matrices $ {\bf A} = (A_1,...,A_k) $:
 \beqn
 {\bf A}_{X,Y} = (XA_{1}Y,...,XA_{k}Y) , \det(XY)=\det(X) \det(Y) = 1
 \eeqn
 As
 $$
  P_{{\bf A}_{X,Y}}(x_1,...,x_k) =  \det( \sum_{1 \leq i \leq k}  x_{i}XA_{i}Y) = \det(X) \det(Y)  P_{{\bf A}}(x_1,...,x_k)
$$
this group action does not change corresponding determinantal polynomial . \\
Finally , it follows that Wick formula is a particular case of formula(24) when there exist two matrices $X$ and $Y$
such that $\det(XY)=\det(X) \det(Y) = 1$ and matrices $(XA_{1}Y,...,XA_{k}Y) $ are all upper triangular ,
or , in Lie-algebraic language , there exists two nonsingular matrices $X$ and $Y$ such that the Lie algebra
generated by $(XA_{1}Y,...,XA_{k}Y) $  is solvable .  \\
It seems reasonable to predict that formula(24) might be of good use in combinatorics desribed in \cite{zvon} .

\section {Short probabilistic proof of formula (24) }
\subsection{Hilbert space of analytical functions }
Consider a Hilbert space $L_{k,G}$ of analytic functions 
$$
 f(x_1,x_2 , ..., x_k) = \sum_{(r_1,...,r_k) } a_{r_1,...,r_k} x_{1}^{r_1} x_{2}^{r_2}...x_{k}^{r_k} ,
$$
where  the $G$- inner product  is defined as
\beqn
<f,g>_{G} = \sum_{(r_1,...,r_k) } a_{r_1,...,r_k} \overline{b_{r_1,...,r_k}}  r_1!...r_k!
\eeqn
It is easy to see that $L_{k,G}$ is a closed proper subspace of $L_{2}(C^{k}, \mu)$ , where $\mu$ is a gaussian measure on $C^{k}$ ,
i.e. its density function 
$$
p(z) = \frac{1}{\pi^{k}} e^{-|z|^{2} }
$$
 
\pro
Suppose that $f , g  \in L_{2}(C^{k}, \mu) $ and the matrix $U : C^{k} \rightarrow C^{k} $ is unitary ,
i.e. $UU^{*} = I $  .
Then 
$$
<f(U x) , g >_{L_{2}(C^{k}, \mu)} = <f , g(U^{*} x) >_{L_{2}(C^{k}, \mu)}
$$
\epro 
\prf
This is just a reformulation of a well known obvious fact that $ p(z) = p(Uz)$ $ (e{-| z |^{2} } = e{-| Uz |^{2} }) $
for unitary $U$ .
\eprf
 \lem
 Let $P(x_1,x_2 , ..., x_k)$ be a homogeneous polynomial of total degree $N$ and $g \in L_{2}(C^{k}, \mu)$ .
 Then for any matrix $A : C^{k} \rightarrow C^{k} $ the following identity holds :
 \beqn
<P(A x) , g >_{L_{2}(C^{k}, \mu)} = <f , g(A^{*} x ) >_{L_{2}(C^{k}, \mu)}
\eeqn
\elem
\prf
First , there is an unique decomposition $g = Q + \delta$ , where $Q(x_1,x_2 , ..., x_k)$ is a  
homogeneous polynomial of total degree $N$ and $<R , \delta >_{L_{2}(C^{k}, \mu)} = 0$
for any homogeneous polynomial $R$ of total degree $N$ . \\
As $P(A x )$ is a homogeneous polynomial of total degree $N$ for all $A$ thus
$<P(A x) , \delta >_{L_{2}(C^{k}, \mu)} \equiv 0$ . \\
It is left to prove (49) only when $g$ is  a homogeneous polynomial of total degree $N$ .
We already know that (49) holds for unitary $A$ . Also , because of formula(23) ,
in this homogeneous case (49) holds for diagonal $A$ . To finish the proof ,
we use the singular value decomposition $A = VDiag U$ , where $U , V$ are unitary
and $Diag$ is a diagonal matrix with a nonnegative entries .
\eprf 
\rem
A homogeneous part of Lemma has been proved in \cite{reznick} using the fact that
the linear space of homogeneous polynomials of total degree $N$ is spanned by
$N$ powers of linear forms . 
\erem
\subsection{Unbiased estimators for Quantum permanents } 
\rem
Consider a four-dimensional tensor $\rho(i_1,i_2,i_3,i_4) ,   1 \leq i_1,i_2,i_3,i_4 \leq N $ .
One can view it as a block matrix as in (3) , where the blocks are defined by
$$
A_{i_{1},j_{1}} = : \{ \rho(i_{1},i_{2},j_{1},j_{2}) : 1 \leq i_{2},j_{2} \leq N \},
 1 \leq i_{1},j_{1} \leq N 
 $$
 We also can permute indices : $\rho(i_{\pi(1) } , i_{\pi(2) },i_{\pi(3) },i_{\pi(4) }) $  , and get
 another block matrix . The main point is that it follows from formula(16) that a permutation
 of indices does not change the quantum permanent $QP(\rho) $ .  In what follows below
 we will use the following simple and natural trick : permute indices and use 
 mixed discriminants based equivalent formula(15) for $QP(\rho) $ based on the corresponding block structure . \\
 \erem
 The next proposition follows directly from the definition .
 \pro
\begin{enumerate}
\item
Consider a block matrix $\rho$  as in (3) and associate with it the following operator $T : M(N) \rightarrow M(N)$ ,
$T(X) =  \sum_{1 \leq i,j \leq N} X(i,j)A_{i,j}$ . \\
Let $X$ be a random complex zero mean matrix such that $E(|X(i,j)|^{2} \equiv 1$ and for any two
permutations $\tau_1 , \tau_2$ the set of entries \\
$ \{X_{i,j} :  j = \tau_{1}(i)  \mbox { or }   j = \tau_{2}(i)  \}$ consists of independent random variables . \\
Then
\beqn
QP(\rho) = E ( \det (T(X))  \overline{\det (X)} ) .
\eeqn
\item
Consider a $N \times N$ matrix A and a random zero mean vector $z \in C^{N}$ such that
$E(z_{i}  \overline{z_{j}}) = 0$ for all $ i \neq j $ . Then
\beqn
Per(A) = E ( \prod_{1 \leq i \leq N} (Tz)_{i}   \overline { \prod_{1 \leq i \leq N} z_{i} } )
\eeqn
\end{enumerate}
\epro
 Let us present now a promised short probabilistic proof of (24 ) : \\
 Consider without loss of generality a $N^{2}$-tuple of  $N \times N$ complex matrices $ {\bf A} = (A_{(1,1)},...,A_{(N,N)}) $. 
 Recall that  the ({\bf BUDM })   $\rho_{{\bf A}}  $   associated with the
$k$-tuple \\ 
$ {\bf A} = (A_1,...,A_k) $ is defined as :
$$
\rho_{{\bf A}}(i_{1},i_{2},j_{1},j_{2}) = : \sum _{1 \leq l \leq k}  A_{l}(i_{1},i_{2}) \overline{A_{l}(j_{1},j_{2}) }
$$
Define the following operator $T(X) =  \sum_{1 \leq i,j \leq N} X(i,j)B_{i,j}$  ,
where $N \times N$ matrix $B_{i,j} = \{\rho_{{\bf A}}(i,j,m,l) : 1 \leq m,l \leq N \} $ . 
It is easy to see that  $T(X) = C^{*}C  (X) $ , where $C(X) = \sum_{1 \leq i,j \leq N} X(i,j)A_{(i,j)}$ .
Thus $QP(\rho_{{\bf A}}) = E_{X} (\det(T(X) \overline {\det(X)} )$ , where  random gaussian matrix
$X$ has the density
$$
p(X) = \frac{1}{\pi^{N^{2}}} e^{-(tr(XX^{*}) }.
$$
I.e. the entries $X(i,j)$ are IID canonical complex gaussian random variables . \\
Finally , we get that (using for the first identity (50) and for the second (49) ) 
$$
QP(\rho_{{\bf A}}) = E_{X} (\det(T(X) \overline {\det(X)} ) = E_{X} (|\det(C(X)|^{2})
$$
But , $ E_{X} (|\det(C(X)|^{2}) = || P_{{\bf A}}||_{G}^{2} $  from (23). \\

Similarly , the permanental formula (43) can be proved using (51) .

\end{document}